\documentclass[prl,preprint,onecolumn,superscriptaddress]{revtex4}

\usepackage{bm}
\usepackage{graphics}
\usepackage{amsmath,epsfig}
\usepackage{amssymb}
\usepackage{comment}
\begin{document}

\title{Quantum-Relay-Assisted Key Distribution over High Photon Loss Channels}

\author{An-Ning Zhang}
\affiliation{Department of Modern Physics and Hefei National
Laboratory for Physical Sciences at Microscale, University of
Science and Technology of China, Hefei, Anhui 230026, People's
Republic of China}
\author{Yu-Ao Chen}
\affiliation{Physikalisches Institut, Universit\"{a}t Heidelberg,
Philosophenweg 12, 69120 Heidelberg, Germany}
\author{Chao-Yang Lu}
\affiliation{Department of Modern Physics and Hefei National
Laboratory for Physical Sciences at Microscale, University of
Science and Technology of China, Hefei, Anhui 230026, People's
Republic of China}
\author{Xiao-Qi Zhou}
\affiliation{Department of Modern Physics and Hefei National
Laboratory for Physical Sciences at Microscale, University of
Science and Technology of China, Hefei, Anhui 230026, People's
Republic of China}
\author{Zhi Zhao}
\affiliation{Physikalisches Institut, Universit\"{a}t Heidelberg,
Philosophenweg 12, 69120 Heidelberg, Germany}
\author{Qiang Zhang}
\affiliation{Department of Modern Physics and Hefei National
Laboratory for Physical Sciences at Microscale, University of
Science and Technology of China, Hefei, Anhui 230026, People's
Republic of China}
\author{Tao Yang}
\affiliation{Department of Modern Physics and Hefei National
Laboratory for Physical Sciences at Microscale, University of
Science and Technology of China, Hefei, Anhui 230026, People's
Republic of China}
\author{Jian-Wei Pan}
\affiliation{Department of Modern Physics and Hefei National
Laboratory for Physical Sciences at Microscale, University of
Science and Technology of China, Hefei, Anhui 230026, People's
Republic of China} \affiliation{Physikalisches Institut,
Universit\"{a}t Heidelberg, Philosophenweg 12, 69120 Heidelberg,
Germany}

\date{\today}

\begin{abstract}
The maximum distance of quantum communication is limited due to
the photon loss and detector noise. Exploiting entanglement
swapping, quantum relay could offer ways to extend the achievable
distance by increasing the signal to noise ratio. In this letter
we present an experimental simulation of long distance quantum
communication, in which the superiority of quantum relay is
demonstrated. Assisted by quantum relay, we greatly extend the
distance limit of unconditional secure quantum communication.
\end{abstract}

\maketitle

In recent years, most of the significant experimental advances
achieved in the field of quantum communication (QC) are based on
the use of photonic channels. However, serious problems occur in
long distance case. Purification \cite{bennett96, pan2001} has
been proposed to regain a high degree of entanglement which is
normally decreases exponentially with the length of the connecting
channel, owing to unavoidable decoherence in the QC channel.
However, more serious problem is caused by combination of
exponential losses of the photons and the dark count of the
detectors which limits today's fiber-based QC systems to operation
over the order of 100 kilometers \cite{gisin02rmp, distance}. The
losses by themselves only reduce the bit rate which is also
exponential with distance. With perfect detectors the distance
would not be limited. However, because of the dark counts, each
time a photon is lost there is a chance that a dark count produces
an error. Hence, when the probability of a dark count becomes
comparable to the probability that a photon is correctly detected,
the signal-to-noise ratio (SNR) tends to zero.

Quantum repeaters \cite{briegel} that combines entanglement
purification \cite{bennett96, pan2001}, entanglement swapping
\cite{zukowski93prl} and quantum memory have been proposed to
overcome these difficulties. Although significant experimental
advances in this direction \cite{pan03puri, Kuzmich03etal,
repeater-zhao-japan} have been achieved, a full quantum repeater
has not yet been realized with current technology. Fortunately, a
much simpler scheme comes as a surprise, which is now commonly
called as a ``quantum relay'' \cite{gisin02rmp, jacobs02}. By
making use of entangled photons and entanglement swapping, it
works similarly as a quantum repeater, but is much closer to be
implemented since it does not need the ability to store photons.
Although a quantum relay does not avoid the exponential loss of
signal in the communication channel, it does increase the SNR and
thus can be used to extend the communication distance.

Previous work along this line includes theoretical investigations
of why and how entanglement swapping helps enhance the maximum
range of QC \cite{ gisin02rmp, yamamoto02, jacobs02,
collins03quantph}, a proof-of-principle experimental demonstration
of entanglement swapping \cite{pan98swap}, an experimental
long-distance (2.2km, optical fiber) entanglement swapping in a
quantum relay configuration \cite{gisin04prl}, in which the main
effort was devoted to preserve the indistinguishablility of the
two photons involved in the Bell state measurement (BSM), and more
recently a long distance (2.2km, optical fiber) experimental
entanglement swapping with a visibility high enough to violate a
Bell inequality \cite{gisin04quantph}. Impressive as it is, there
is no experimental work to study the effect of quantum relay yet.
It would be highly desirable to see in a practical way whether or
not and, to what extent, quantum relay will show its advantage on
extending the available distance of QC.

In this letter, we report an experimental simulation of long
distance quantum communication based on quantum relay. The
interest of quantum relay is shown by performing quantum key
distribution (QKD) over a high photon loss channel that a secure
BB84 protocol \cite{eprBB84} cannot be realized. To do QKD over a
longer distance with the current technology, two-way classical
communication could be a good choice \cite{Lo,Chou,wang}. The
threshold of quantum bit error rate (QBER) of unconditional secure
BB84 protocol extends to $20.0\%$ \cite{Chou} by two-way classical
communication. In our experiment, the superiority of long-distance
distribution of entangled photons assisted with entanglement
swapping over the way of direct transmission is shown by extending
the absolutely secure channel length to about two times.

Let us first review the salient feature of the quantum relay.
Without going into complicated calculations
\cite{collins03quantph}, we only give the essence and main results
here. As shown in Fig. 1(a), the simplest way to distribute an
entangled state is to directly send a photon from a sender to a
receiver, say Alice and Bob. At short distance, the visibility is
high because dark counts are negligible. However, as the distance
increases, more and more photons will be absorbed while the dark
counts will remain constant, giving a lower SNR and thus the
maximal distance for a given fidelity is limited. The
implementation of a quantum relay using quantum nondemolition
(QND) measurement is illustrated in Fig. 1(b). The basic idea
about quantum relay is to verify at some point along the channel
whether the photon is still here \cite{jacobs02}. If the photon is
still present, a classical signal indicating this fact is sent to
Bob. If a photon is not detected, which means that the correct
photon has been lost in its way, and then a classical signal tells
Bob not to accept any output from his detectors. As a result, the
limiting effect of the detector dark count can be essentially
suppressed.

Using the simple linear optical device, we could conveniently
implement a probabilistic QND measurement by adopting the scheme
of quantum teleportation \cite{dik97, dowling02pra} as shown in
Fig. 1(c). Resources needed are the maximally entangled photon
pairs and the polarizing beam splitter (PBS), which serves as a
probabilistic Bell state analyzer \cite{pan98pra}. It is easily
seen that if any one of the two incoming photons is absent, a
coincidence of the two detectors behind the PBS never happen.
However, if both of them are still present, BSM will
probabilistically succeed and faithfully transfer the quantum
state onto the photon towards Bob. Taking into account of
realistic parameters including the loss of the channel, the
detector dark counts etc., numerical simulations on the
performance of quantum relay have been presented in Ref.
\cite{jacobs02, collins03quantph} to show the advantage of using
quantum relay.

\begin{figure}[t]
  \begin{center}
  \includegraphics[width=3.1in]{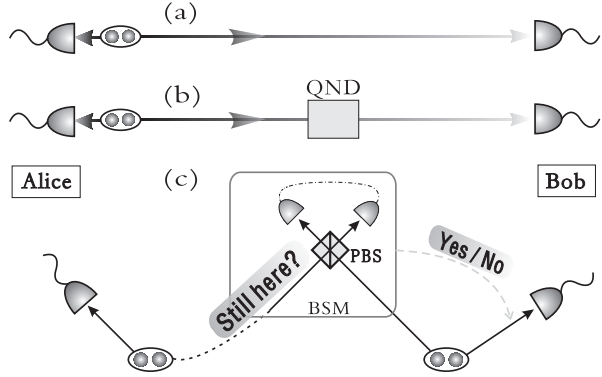}
  \end{center}
  \caption{Graphical show of working principle of quantum relay.
  (a) Entangled state distribution by directly sending a photon.
  (b) Quantum relay works by exploiting QND measurement to verify
  at some point whether the photon is still here.
  (c) Entanglement swapping is employed to implement a probabilistic
  QND measurement.}
    \label{relay}
\end{figure}

\begin{figure}[b]
  \begin{center}
  \includegraphics[width=3.1in]{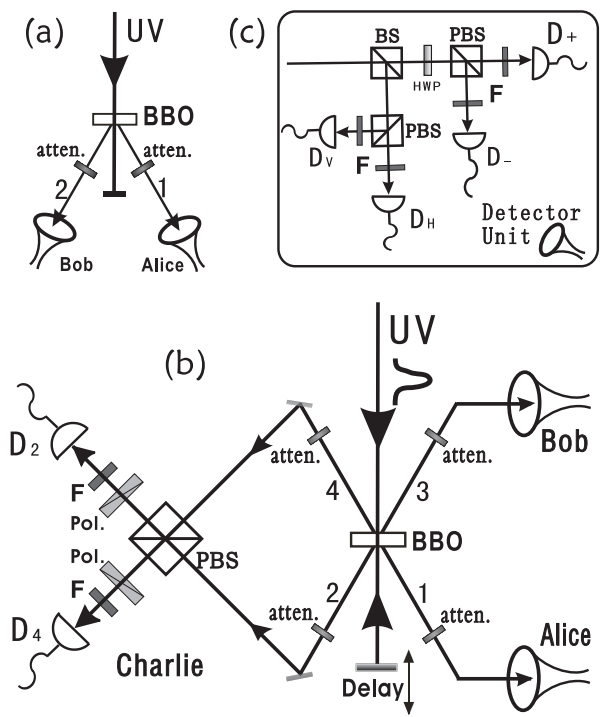}
  \end{center}
  \caption{Experimental setup for simulation of long distance
  quantum cryptography. An ultraviolet pulsed laser from a
  mode-locked Ti:sapphire laser (center wavelength $394nm$,
  pulse duration $200fs$, repetition rate $76MHz$) passed through
  $\beta$-barium borate (BBO) crystal twice to generate two
  maximally entangled photon pairs in modes 1-2 and mode 3-4.
  In the experiment, we managed to obtain an average twofold
  coincidence rate of $2.4\times10^4 s^{-1}$ with an average pump
  power of $470mW$. (a) Entangled photons are distributed by
  direct transmission. Attenuation plate (atten.) is inserted
  to simulate the distance. (b) Entangled state distribution
  assisted with another EPR pair (3-4) and entanglement swapping.
  (c) Photon detector unit for QKD.}
    \label{setup}
\end{figure}

A realistic experimental demonstration of the superiority of
quantum relay will take many efforts, e.g. to distribute photons
over very long distance and to implement successful BSM. Here we
avoid these complexities by performing a simulative experiment. In
our experiment we aim to compare the performance of QKD over high
photon loss channel with and without quantum relay. We examine
under the scenario that the photons are emitted from a perfect
entangled photon source (PEPS), that is, exactly one photon pair
per pulse. Then we equivalently consider the actual probabilistic
photon source produced by parametric down-conversion (PDC) as like
the photons from the PEPS have passed a certain distance in a
lossy channel, during which the photons are probabilistically
lost. Similarly, additional attenuation plates placed into the
light path are also used to simulate the distances. Sum up the
above two effects and we get the total attenuation. By using these
reasonable equivalences, this experiment is greatly simplified but
does not prevent us from showing its physical essence.

In the experiment, we relate the QBER to the SNR and use the QBER
to characterize the performance of a quantum relay. A schematic
drawing of our experimental setup is shown in Fig. 2. The required
photon pairs are produced via Type-II PDC and prepared in the
state $ |\Psi ^{-}\rangle_{ab} = |H\rangle_a |V\rangle_b
-|V\rangle_a |H\rangle_b $ with a high SNR of 30:1 in the $+/-$
basis, where $|\pm\rangle=|H\rangle\pm|V\rangle$ (coefficients
omitted).

In our experiment, we first test the performance of direct
entangled photons distribution without assistance of quantum
relay. Using the experimental setup shown in Fig. 2(a), we employ
entangled photon pair 1-2 to implement quantum key distribution.
We place in the light path a series of attenuation plates with a
transmission rate range from $ t_a = 0.27 $ to $ t_a = 6 \times
10^{-3} $ which, together with the equivalent attenuation effect
from probabilistic PDC photon source $t_s = c/(76M\times \eta^2)$,
constitute the total attenuation $ t=t_s \times t_a^2 $. Here
$c=2.4\times 10^4s^{-1}$ is the two-fold coincidence rate,
$\eta=0.15$ is the average detection efficiency.

In order to experimentally demonstrate the BB84 protocol, both
Alice and Bob need to perform a polarization measurement on their
own photons by randomly choosing $H/V$ or $+/-$ polarization
basis. To achieve the random choice of the measurement basis, we
let the photon pass through a 50-50 beam splitter (BS) as in Fig.
2(c). In one of the two BS outputs a polarizing beam splitter is
used to perform the $H/V$ polarization analysis. In another output
of the BS, a half-wave plate (HWP) is put in front of the PBS and
oriented at 22.5$^{0}$ to measure the photon along the $+/-$
polarization basis. From the experimental results as shown in Fig.
3 (triangle dots), we can see that as the equivalent distance
increases, the QBER arises dramatically. At the attenuation of
about $37.5dB$, the QBER reaches 20.0\%, a security bound of
quantum key distribution \cite{Chou}, indicating that over
attenuation condition, the unconditional security can not be
guaranteed by two-way classical communication.

To accomplish the secure QKD task over the channel attenuation
limit of $37.5dB$, we now take the advantage of quantum relay. In
the experiment, another pair of EPR photons required is produced
via the PDC process by the UV pulse after its reflection, whose
two-fold coincidence rate is adjusted to be less than 5\%
different from the pair 1-2. We then superpose the photons 2 and 4
at the PBS to implement the BSM. Their path lengths are adjusted
such that they arrive simultaneously. To achieve good spatial and
temporal overlap, the outputs are spectrally filtered
$(\triangle\lambda=2.8nm)$ and monitored by fiber-coupled
single-photon detectors. These processes effectively erase any
possibility of distinguishing the two photons and thus lead to
interference. Conditioned on the coincidence of Detector 2 and 4
behind the PBS, the entanglement of photon 2 and 4 is swapped to
the photon 1 and 3 (coefficients omitted):
\begin{eqnarray*}
&&(|H\rangle_{1}|V\rangle _{2}-|V\rangle _{1}|H\rangle _{2}) \otimes (|H\rangle _{3}|V\rangle _{4}-|V\rangle _{3}|H\rangle _{4})  \nonumber \\
&\rightarrow&|H\rangle _{1}|V\rangle _{2}|H\rangle
_{3}|V\rangle _{4}+|V\rangle _{1}|H\rangle _{2}|V\rangle _{3}|H\rangle _{4} \nonumber \\
&=&|\Phi ^{+}\rangle _{13}|\Phi ^{+}\rangle _{24}-|\Phi
^{-}\rangle _{13}|\Phi ^{-}\rangle _{24} \nonumber
\end{eqnarray*}

\begin{figure}[t]
  \begin{center}
  \includegraphics[width=3.5in]{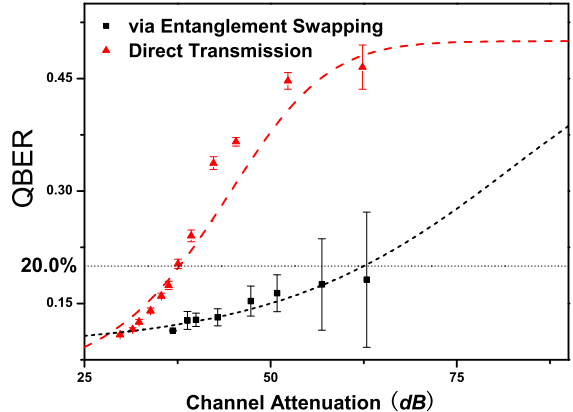}
  \end{center}
  \caption{Experimental results showing the advantage of quantum relay.
  The plots are measured QBERs at a certain equivalent distance.
  The shot dash line and dash line are numerical simulations according to the real
  parameters in our experiment with and without quantum delay, respectively.}
    \label{compare}
\end{figure}

In the BSM, we only register those cases where both D2 and D4
detect a $|+\rangle$ polarized photon for experimental simplicity,
which consequently projects the photons 1 and 3 into the state
$|\Phi^{+}\rangle_{13}$. Similarly we measure the QBER under a set
of attenuation condition, which is also shown in Fig. 3 (square
dots). The total equivalent attenuation rate this time is $ t =
t_s^2 \times t_{a'}^4 $, which consists of two probabilistic PDC
source and four attenuation plates (with a transmission rate
$t_{a'}$ ranging from 1 to 0.22) inserted in the light path as
shown in Fig. 2(b). We can see from the Fig. 3 that assisted with
quantum relay the QBER arises significantly slower, demonstrating
its superiority. The attenuation where the QBER reaches 20.0\%
this time is about $65dB$.

Taking into account of the realistic experimental parameters: a
success probability of $1/8$ implementing BSM, an average
probability of having a dark count in any single-photon detector
$D = 1.1 \times 10^{-4}$ (see \cite{note} for definition), the
average overall collection and detection efficiency $\eta = 0.15$,
we follow the method developed in ref. \cite{collins03quantph} to
carry out a numerical simulation [See Appendix] of the performance
of our relay system, which is shown in Fig. 3 (dash lines). It can
be seen from Fig. 3 that the experimental data shows a good
agreement with the numerical simulation. Moreover, for the sake of
clarity, we can transform the total attenuation into the
equivalent distance of the communication channel with a loss of
$\alpha = 0.25dB/km$, the typical photon loss in the telecom fiber
channel. Thus, it is clearly shown that quantum relay extend the
secure distance of quantum key distribution from $150km$ up to
$260km$.

In conclusion, by reasonable equivalences, we simulate an
experiment of long distance quantum key distribution. We
demonstrate the advantage of long-distance distribution of
entangled photons assisted with quantum relay over the method of
direct transmission. For the first time we experimentally show
that the distance of QC could be extended by making use of
entanglement swapping, which is quite simple but powerful. Since
entanglement swapping also serves as a key element in the scheme
of quantum repeater, the experimental methods develop here may
also be helpful for the future realization of quantum repeaters.
Furthermore, by proper modifications, this experimental scheme
allows an immediate experimental demonstration of third-man
quantum cryptography and quantum secret sharing \cite{thirdman,
sharing}. While further practical investigations on real world are
still necessary, we believe that this work provides a useful
toolbox for tomorrow's long distance realization of QC.

We acknowledge J. R\"osch for helpful comments and X.-H. Bao, J.
Zhang and C.-Z. Peng for technical supports. This work was
supported by the NNSF of China, the Chinese Academy of Sciences.
This work was also supported by the Marie Curie Excellent Grant of
the EU, the Alexander von Humboldt Foundation and the Deutsche
Telekom Stiftung.

\newpage
{\bf Appendix: Analysis of QBER in Quantum-Relay-Assisted Key
Distribution}
\\

As has been mentioned in the paper, in our experiment we aim to
test the performance of quantum key distribution (QKD) over high
photon loss channel with and without quantum relay. We examine
under the scenario that the photons are emitted from a perfect
entangled photon source (PEPS), that is, exactly one photon pair
per pulse. Then we equivalently consider the actual probabilistic
photon source produced by parametric down-conversion (PDC) as like
the photons from the PEPS have passed a certain distance in a
lossy channel, during which the photons are probabilistically
lost. Similarly, additional attenuation plates placed into the
light path are also used to simulate the distances. Sum up the
above two effects and we get the total attenuation ($t$).

In the experiment we use an ultraviolet pulsed laser from a
mode-locked Ti:sapphire laser (pump power 470mW, center wavelength
394nm, pulse duration 200fs, repetition rate 76MHz). And we
equivalently consider the entangled photon pair produced by PDC as
the PEPS emitted with a frequency of $76M/s$ passing through a
quantum channel whose transmission probability is $t_{s}$. We
observed in the experiment a two fold coincidence:
$c=2.2\times10^{4}$ and an average detector efficiency
$\eta=0.15$. And thus $t_{s}$ is
\begin{equation}
t_{s}=\frac{c/\eta^2}{76M}\label{epr}
\end{equation}

Additional attenuation plates placed into the light path are also
used to simulate the distances. The transmission probability of
the channel with attenuation plates is $t_{c}$. The total
transmission probability $t$ is $t=t_{c}\times t_{s}$.

In our experiment, we measured an average count of $20000s^{-1}$
under a moderate ambient lighting. The measurement time window is
set to be $5.5ns$ so the probability of having a dark count
probability in one detector is: $D=1.1 \times 10^{-4}$.

Following the method presented in Ref. \cite{collins03quantph}, we
perform a numerical simulation as follows:

First let us consider the case of direct QKD. For Alice and Bob to
correctly receive and accept the signal, the photons must pass the
channel unabsorbed ($t$), and be detected by the detector
($\eta$). So, the probability of existing an output at Alice's
(Bob's) detectors in the time window is
\begin{equation}
P(Alice)=P(Bob)=[t^{\frac{1}{2}}\eta+(1-t^{\frac{1}{2}}\eta)4D](1-D)^{3}.
\end{equation}
Here, in our experiment, we perform the symmetrical attenuation in
the two channel of Alice and Bob, therefore, the attenuation of
the each channel is $t^{\frac{1}{2}}$.

Thus the probability of sifted key bit is
\begin{eqnarray}
P(total)&=&\frac{1}{2}\cdot P(Alice) \cdot P(Bob) \nonumber \\
&=&\frac{1}{2}\{[t^{\frac{1}{2}}\eta+(1-t^{\frac{1}{2}}\eta)4D](1-D)^{3}\}^{2}.
\end{eqnarray}
Because the unperfect futures of the source and the channel, in
our experiment, we observe the optical visibility ($V_{opt}$) of
single photon is $0.95$. Therefore, the probability of creating a
correct key (signal) is,
\begin{equation}
P(signal)=\frac{1}{2}V_{2opt}[t^{\frac{1}{2}}\eta (1-D)^{3}]^{2},
\end{equation}
$V_{2opt}=0.95^{2}$ is the optical visibility of the photon pair.

When we perform the QKD via quantum relay, we perform the same
attenuation in the four photon channel, therefore, the attenuation
of the each channel is $t^{\frac{1}{4}}$. Similarly, in the time
window the probability of Alice and Bob is
\begin{equation}
P_{QR}(Alice)=P_{QR}(Bob)=[t^{\frac{1}{4}}\eta+(1-t^{\frac{1}{4}}\eta)4D](1-D)^{3}.
\end{equation}

In our experiment, the polarized beam-splitter and two
single-photon detectors behind two polarizers setting at $45^{0}$
constitute the Bell-state measurement (BSM) analyzer. The
probability of that there is a BSM output $P(Bell)$ is
\begin{equation}
P(Bell)=\frac{1}{2}(t^{\frac{1}{4}}\eta\eta_{p})^{2}+[(1-t^{\frac{1}{4}}\eta\eta_{p})D]^{2}+2t^{\frac{1}{4}}\eta\eta_{p}(1-t^{\frac{1}{4}}\eta\eta_{p})D,
\end{equation}
here $\eta_{p}=0.5$ denotes the transmission probability of the
polarizer in the BSM.
$\frac{1}{2}(t^{\frac{1}{4}}\eta\eta_{p})^{2}$ is the probability
of detecting one photon in each detector;
$[(1-t^{\frac{1}{4}}\eta\eta_{p})D]^{2}$ is the contribution from
the dark count in both detectors; the probability of registering
one photon in a detector while dark count in another is
$2t^{\frac{1}{4}}\eta\eta_{p}(1-t^{\frac{1}{4}}\eta\eta_{p})D$.
\\

Thus the probability of sifted key bit is
\begin{eqnarray}
P_{QR}(total)&=&P(Bell) \cdot \frac{1}{2} \cdot P_{QR}(Alice)
\cdot P_{QR}(Bob)
\nonumber\\
&=&\{\frac{1}{2}(t^{\frac{1}{4}}\eta\eta_{p})^{2}+[(1-t^{\frac{1}{4}}\eta\eta_{p})D]^{2}
+2t^{\frac{1}{4}}\eta\eta_{p}(1-t^{\frac{1}{4}}\eta\eta_{p})D\}
\cdot\nonumber\\
&&\frac{1}{2}[t^{\frac{1}{4}}\eta+(1-t^{\frac{1}{4}}\eta)4D]^{2}(1-D)^{6}.
\end{eqnarray}

The signal is
\begin{equation}
P_{QR}(signal)=V_{4opt}\{\frac{1}{2}[t^{\frac{1}{4}}\eta
(1-D)^{3}]^{2}\}\cdot \lbrack
\frac{1}{2}(t^{\frac{1}{4}}\eta\eta_{p} )^{2}],
\end{equation}
where $V_{4opt}=0.95^{4}$.

$QBER$ of QKD is
\begin{equation}
QBER=\frac{1}{2}[1-\frac{P(signal)}{P(total)}].
\end{equation}

When we perform the QKD via transmitting the entangled photon
pairs directly,
\begin{equation}
QBER=\frac{1}{2}[1-\frac{V_{2opt}[t^{\frac{1}{2}}\eta]^{2}}{{[t^{\frac{1}{2}}\eta+(1-t^{\frac{1}{2}}\eta)4D]}^{2}}].
\end{equation}

When we perform the QKD via quantum relay,
\begin{eqnarray}
& QBER_{QR} \ \ \ \ \ \ \ \ \ \ \ \ \ \ \ \ \ \ \ \ \ \ \ \ \ \ \ \ \ \ \ \ \ \ \ \ \ \ \ \ \ \ \ \ \ \ \ \ \ \ \ \ \ \ \ \ \ \ \ \ \ \ \ \ \ \ \ \ \ \ \ \ \ \ \ \ \ \ \ \ \nonumber\\
&=\frac{1}{2}[1-\frac{V_{4opt}(t^{\frac{1}{4}}\eta
)^{4}\eta_{p}^{2}}
{2\{\frac{1}{2}(t^{\frac{1}{4}}\eta\eta_{p})^{2}+[(1-t^{\frac{1}{4}}\eta\eta_{p})D]^{2}
+2t^{\frac{1}{4}}\eta\eta_{p}(1-t^{\frac{1}{4}}\eta\eta_{p})D\}
\cdot
[t^{\frac{1}{4}}\eta\eta_{p}+(1-t^{\frac{1}{4}}\eta)4D]^{2}}].
\end{eqnarray}


\begin{thebibliography}{00}
\bibitem{note} The parameter $D$ is defined as the probability of having a
dark count in any single-photon detector within our experiemental
coincidence window of $5.5ns$. In the experiement, with the
entangled photon source blocked, an average single count rate of
$20000s^{-1}$ are observed under a moderate ambient lighting, this
leads to a relative high $D = 1.1 \times 10^{-4}$.
\end{thebibliography}
\end{document}